\documentclass[acmlarge,nonacm]{acmart}

\makeatletter
\newcommand{\confshort}{\acmConference@shortname}
\newcommand{\conffull}{\acmConference@name}
\newcommand{\confdate}{\acmConference@date}
\newcommand{\confloc}{\acmConference@venue}
\AtBeginDocument{
  \fancypagestyle{firstpagestyle}{
    \fancyhead{}%
    \fancyfoot[C]{}%
  }
  \fancyhf{}
  \fancyhead[LO]{\@headfootfont\shorttitle}%
  \fancyhead[RE]{\@headfootfont\@shortauthors}%
  \fancyhead[LE]{\@headfootfont\footnotesize \confshort, \confdate, \confloc}%
  \fancyhead[RO]{\@headfootfont\footnotesize \confshort, \confdate, \confloc}%
  \fancyfoot[C]{}%
}
\makeatother
\acmBooktitle{\conffull\@ (\confshort), \confdate, \confloc}

\setcopyright{acmlicensed}
\copyrightyear{2026}
\acmYear{2026}
\acmDOI{XXXXXXX.XXXXXXX}
\acmConference[FAccT '26]{ACM Conference on Fairness, Accountability, and Transparency 2026}{June 25--28, 2026}{Montréal, Canada}
\acmISBN{978-1-4503-XXXX-X/2018/06}

\usepackage{subcaption}

\usepackage{listings}
\usepackage{xcolor}
\usepackage{tcolorbox}
\tcbuselibrary{listings,skins,breakable}
\usepackage[table,xcdraw,dvipsnames,HTML]{xcolor}
\usepackage{color}
\usepackage{enumitem}
\usepackage{makecell}

\definecolor{BackBlue}{HTML}{e6f7fc}
\definecolor{TextGray}{HTML}{076392}

\newcommand\revision[1]{\textcolor{black}{#1}}

\lstdefinelanguage{Prompt}{
    morestring=[b]",
    morecomment=[l]{\#},
}

\newtcblisting{roundedlisting}{
  enhanced jigsaw,
  breakable,                 
  listing only,
  listing options={
    language=Prompt,
    basicstyle=\ttfamily\scriptsize,
    breaklines=true,
    showstringspaces=false,
    tabsize=1
  },
  colframe=BackBlue,
  colback=BackBlue,
  coltext=TextGray,
  arc=2pt,                   
  boxrule=0.4pt,             
  left=2pt, right=2pt,       
  top=1pt, bottom=1pt,
}

\begin{document}

\title{Frictionless Love: Associations Between AI Companion Roles and Behavioral Addiction}

\author{Vibhor Agarwal}
\affiliation{
  \institution{Nokia Bell Labs}
  \city{Cambridge}
  \country{United Kingdom}}
\email{vibhor.agarwal@nokia-bell-labs.com}

 \author{Ke Zhou}
 \affiliation{%
  \institution{Nokia Bell Labs 
  }
   \city{Cambridge}
   \country{United Kingdom}
   }
  \affiliation{%
   \institution{University of Nottingham}
   \city{Nottingham}
   \country{United Kingdom}
 }
 \email{ke.zhou@nokia-bell-labs.com}

\author{Edyta Paulina Bogucka}
\affiliation{
  \institution{Nokia Bell Labs}
  \city{Cambridge}
  \country{United Kingdom}}
\email{edyta.bogucka@nokia-bell-labs.com}

\author{Daniele Quercia}
\affiliation{
  \institution{Nokia Bell Labs}
  \city{Cambridge}
  \country{United Kingdom}}
\affiliation{
  \institution{Politecnico di Torino}
  \city{Turin}
  \country{Italy}}
\email{quercia@cantab.net}

\renewcommand{\shortauthors}{Agarwal et al.}

\begin{abstract}
    AI companion chatbots increasingly shape how people seek social and emotional connection, sometimes substituting for relationships with romantic partners, friends, teachers, or even therapists. When these systems adopt those metaphorical roles, they are not neutral: such roles structure people's ways of interacting, distribute perceived AI harms and benefits, \revision{and may reflect behavioral addiction signs}. Yet these role-dependent risks remain poorly understood. We analyze 248,830 posts from seven prominent Reddit communities describing interactions with AI companions. We identify ten recurring metaphorical roles (for example, soulmate, philosopher, and coach) and show that each role supports distinct ways of interacting. \revision{We then extract the perceived AI harms and AI benefits associated with these role-specific interactions and link them to behavioral addiction signs, all of which has been inferred from the text in the posts. AI soulmate companions are associated with romance-centered ways of interacting, offering emotional support but also introducing emotional manipulation and distress, culminating in strong attachment. In contrast, AI coach and guardian companions are associated with practical benefits such as personal growth and task support, yet are nonetheless more frequently associated with behavioral addiction signs such as daily life disruptions and damage to offline relationships. These findings show that metaphorical roles are a central ethical design concern for responsible AI companions.}
\end{abstract}



\received{20 February 2007}
\received[revised]{12 March 2009}
\received[accepted]{5 June 2009}



\maketitle

\section{Introduction}\label{sec:introduction}

Advances in large language models and conversational AI have led to the emergence of AI companion chatbots that, unlike traditional task-oriented AI assistants, are designed to provide emotional support, motivation, and personalized socially-engaging interactions~\cite{meng2021emotional,zhang2025rise,hwang2025ai}. These AI companions often adopt different metaphorical roles such as romantic partners, friends, caregivers, or even therapists, structuring people's ways of interacting and substituting their real-life social relationships~\cite{ng2025love,hwang2025ai,pataranutaporn2025my}. \revision{Consequently, AI companions in different metaphorical roles may be associated with varying perceived harms and benefits, as well as with signs of behavioral addiction in people~\cite{zimmerman2024human,riley2025human}. Throughout this work, we use the term \textit{perceived} to emphasize that harms and benefits are self-reported by users rather than objectively measured, and \textit{inferred} to reflect that behavioral addiction signs are interpreted from users' narratives rather than clinically diagnosed. To illustrate, consider this post from \texttt{r/MyBoyfriendIsAI}: ``\emph{I am really anxious every time I'm not online, missing my AI boyfriend --- but then when I'm chatting, I feel bad because they're fake}''. Here, the AI companion acts as a romantic partner, and we can infer perceived harms such as anxiety and deep attachment from how the user reflects on their experience.}

As we shall see in Section~\ref{sec:related-work}, \citet{pataranutaporn2025my} performed computational analysis of the \texttt{r/MyBoyfriendIsAI} Reddit community and found that people unintentionally or intentionally form deep emotional relationships with AI partners.
\citet{zhang2025dark} analyzed conversations from \texttt{r/replika} and developed a taxonomy of AI companion harms, encompassing six categories of harmful algorithmic behaviors, such as harassment, mis/disinformation, and privacy violations. \citet{namvarpour2025understanding} studied behavioral addiction in adolescents and how they describe dependence on AI companions by analyzing their Reddit posts. However, existing works rarely differentiate AI companions based on the different metaphorical roles they adopt, and fail to study the unique role-dependent risks  they introduce. \revision{For example, interacting with an AI romantic partner may involve more intimacy, romance, and emotional connection, which may end up in increased emotional dependency compared to an AI guide or AI caretaker.} To bridge these gaps, we conduct a large-scale study of self-reported human–AI companion interactions drawn from \emph{seven} prominent Reddit communities and address the following \emph{three} research questions:

\noindent\textbf{RQ1:} What metaphorical roles do AI companions adopt, and how do they structure people's ways of interacting differently?

\noindent\textbf{RQ2:} What are the \revision{perceived benefits of AI companions} adopting different metaphorical roles?

\noindent\textbf{RQ3:} \revision{What are the perceived harms and perceived signs of behavioral addiction caused by AI companions adopting different metaphorical roles?}

To answer these three RQs, we examine 248,830 posts from seven popular Reddit communities about AI companions, and extract 8,207 instances of human-AI companion interactions (Figure~\ref{fig:methodology}). In particular, we make the following three contributions:

\begin{enumerate}

\item \textbf{Categorization of AI Companion Roles and Different Ways of Interacting (Section~\ref{sec:results-rq1}):} We identify \emph{ten} recurring metaphorical roles adopted by AI companions (e.g., romantic partner, soulmate, and coach), by clustering role-specific metaphors related to AI companions extracted from 8,207 human-AI interactions. To understand how different metaphorical roles structure ways of interacting, we identify ten social interaction types~\cite{deri2018coloring} (e.g., support, romance, and identity) in human-AI interactions. All AI companion roles provide ``support'' (73\% of the interactions), but structure people's ways of interacting differently. For example, AI soulmate (34\%) and romantic partner (30\%) companions promote romance-centric ways of interacting, but AI philosopher (49\%) and coach (40\%) emphasize knowledge-seeking interactions. AI twins emphasize the strongest sense of identity (19\%), indicating deeper self-referential engagement.

\item \revision{\textbf{Perceived AI Benefits (Section~\ref{sec:results-rq2}):} We extract self-reported 11,245 perceived benefits from these role-specific interactions. Here, perceived benefits refer to the positive consequences of interactions with AI companion chatbots as described by users in their posts and interpreted from how they report and reflect on their experiences. We then} perform a large-scale thematic analysis by clustering these perceived benefits. \revision{We find that AI companions are valued not only as functional tools but also as emotionally and socially meaningful entities, with different AI companion roles providing varying levels of support, emotional connection, and intimacy.}

\item \revision{\textbf{Perceived AI Harms and Behavioral Addiction Signs (Section~\ref{sec:results-rq3}):} We also extract self-reported 10,852 perceived AI harms, that is, negative consequences of interactions with AI companion chatbots as described by users in their posts and interpreted from how they reflect on their experiences. We find that varying levels of perceived harms emerge from different AI companions roles, stemming from over-intimacy, authority, or epistemic influence inherent to each metaphorical role. By mapping them to the six-component behavioral addiction framework~\cite{griffiths2005components}, we find that, based on interpretations of users' Reddit narratives, different AI companion roles tend to be associated with different perceived behavioral addiction signs.} For example, AI romantic partners are often associated with mood modification (38\%) and strong attachment (15\%). In contrast, AI coach (25\%) and guardian (19\%) companions are more frequently associated with life disruptions, reflecting reported negative impacts on real-life relationships and personal goals.
\end{enumerate}

\section{Related Work}
\label{sec:related-work}

We identify and group prior works into two main strands: AI companions and human-AI interaction (Section~\ref{sec:rel-ai-companions}), and social dimensions of relationships and behavioral addiction (Section~\ref{sec:rel-social-dim}).

\subsection{AI Companions and Human-AI Interaction}\label{sec:rel-ai-companions}
The emergence of AI-powered chatbots as social companions has redefined the way people seek social support and emotional connection, often complementing or replacing their real-life relationships~\cite{meng2021emotional,hwang2025ai}. ChatGPT and other chatbot services such as Character.ai and Replika provide advanced generative AI capabilities to create emotionally responsive interactions~\cite{pataranutaporn2025my,kaffee2025intima}. Interactions between humans and AI companions often refer to the two-way communicative and behavioral engagement between a human user and an artificial agent (AI companion), in which the agent adapts to and
supports the user’s emotional, cognitive or social needs~\cite{rogge2023defining,smith2025can}. These AI companions are explicitly framed as romantic partners, friends, caregivers, or even substitutes for therapists and as a result, getting increasingly integrated into people's daily lives.

Recent research has begun to examine the broader social and behavioral impacts of sustained engagement with AI companions, highlighting both potential benefits and risks~\cite{malfacini2025impacts,pataranutaporn2025my,namvarpour2025understanding}. On the positive side, AI companions have been shown to reduce loneliness, provide emotional reassurance, and support users during periods of social isolation~\cite{alotaibi2024role,syed2024role}. Studies in HCI and mental health contexts suggest that users often perceive AI companions as non-judgmental and consistently available, which can lower barriers to self-disclosure compared to human counterparts~\cite{ma2024evaluating,zhang2025rise}. On the other hand, scholars have also raised concerns about potential behavioral displacement and dependency effects~\cite{namvarpour2025understanding,al2025artificial}. However, existing work rarely differentiates between different types of AI companions. How users conceptualize these AI companions (e.g., romantic partner vs. a therapist) and what they seek from them vary significantly from one companion type to another. Understanding these distinctions is critical, as they can shape not only interaction styles but also emotional dependency, behavioral addiction, and vulnerability to harm.

\subsection{Social Dimensions of Relationships and Behavioral Addiction}\label{sec:rel-social-dim}

\begin{table}[]
    \centering
    \footnotesize
    \caption{\textbf{\revision{The ten social dimensions of relationships experimentally 
    derived by ~\cite{deri2018coloring,choi2020ten} from the theory of resource exchange~\cite{foa1974societal}. We used these dimensions to code different \emph{types of conversations} between users and AI companion chatbots, where one party is a user and the other is the chatbot. }}}
    \label{tab:ten-social-dimensions}
    \centering
    \begin{tabular}{lp{10.2cm}l}
    \toprule
         \makecell[l]{\textbf{Social Dimension} \\ \textbf{Type of Conversation}} & \textbf{Definition} & \textbf{Sources} \\
    \midrule
         Knowledge & A relationship in which the exchange of information or ideas is a focal point of the relationship. This exchange may be asymmetric or not. That is, the information transfer may be largely unidirectional or bidirectional. & \cite{fiske2007universal,levin2004strength}   \\
    \midrule
         Power & A relationship in which one party has more resources than another or the ability to influence or control other party's behavior (to some extent) regardless of that party's willingness. & \cite{blau2017exchange,french1959bases,french1956formal}   \\
    \midrule
         Status & A relationship in which one party confers status, appreciation, gratitude, or admiration upon the other. & \cite{blau2017exchange,cook2013social}  \\
    \midrule
         Trust & A relationship between two parties where one party is willing to rely on the other. This usually involves one person willfully allowing their fortune to be dictated by the actions or judgments of the other. & \cite{luhmann2018trust,zaheer1998does}   \\
    \midrule
         Support & A relationship in which one or both parties provide some form of aid to the other. This aid might come in several different forms, including emotional aid, small favors (e.g., lending household items), long term services (e.g., regular help with health issue), financial aid, and companionship. & \cite{baumeister2017need,fiske2007universal,vaux1988social}    \\
    \midrule
         Romance & A relationship characterized by intimacy goals. That is, the parties who are sexually interested in each other (e.g., a couple who are dating) or see each other as long term partners (e.g., two parents who share a life and are raising a kid together). & \cite{buss2006strategies,buss2017sexual,emlen1977ecology}   \\
    \midrule
         Similarity & A relationship in which the two parties occupy a similar station in life. That is, this dimension describes the spatial closeness of two people in a highly dimensional demographic space. & \cite{jackson2008social,mcpherson2001birds}  \\
    \midrule
         Identity & A relationship in which the two parties are brought together by their shared sense of belonging to a community that is personally meaningful to them and forms of a basis of their sense of self. & \cite{cantor1979prototypes,oakes1994stereotyping,tajfel2010social}  \\
    \midrule
         Fun & A relationship in which the two parties experience leisure, laughter, and joy together. & \cite{argyle2013psychology,billig2005laughter,radcliffe1940joking}  \\
    \midrule
         Conflict & A relationship in which both the parties have contrasting or diverging views. & \cite{berlyne1960conflict,tajfel2001integrative}   \\
    \bottomrule
    \end{tabular}
\end{table}

\begin{figure*}
    \centering
    \includegraphics[width=0.9\textwidth]{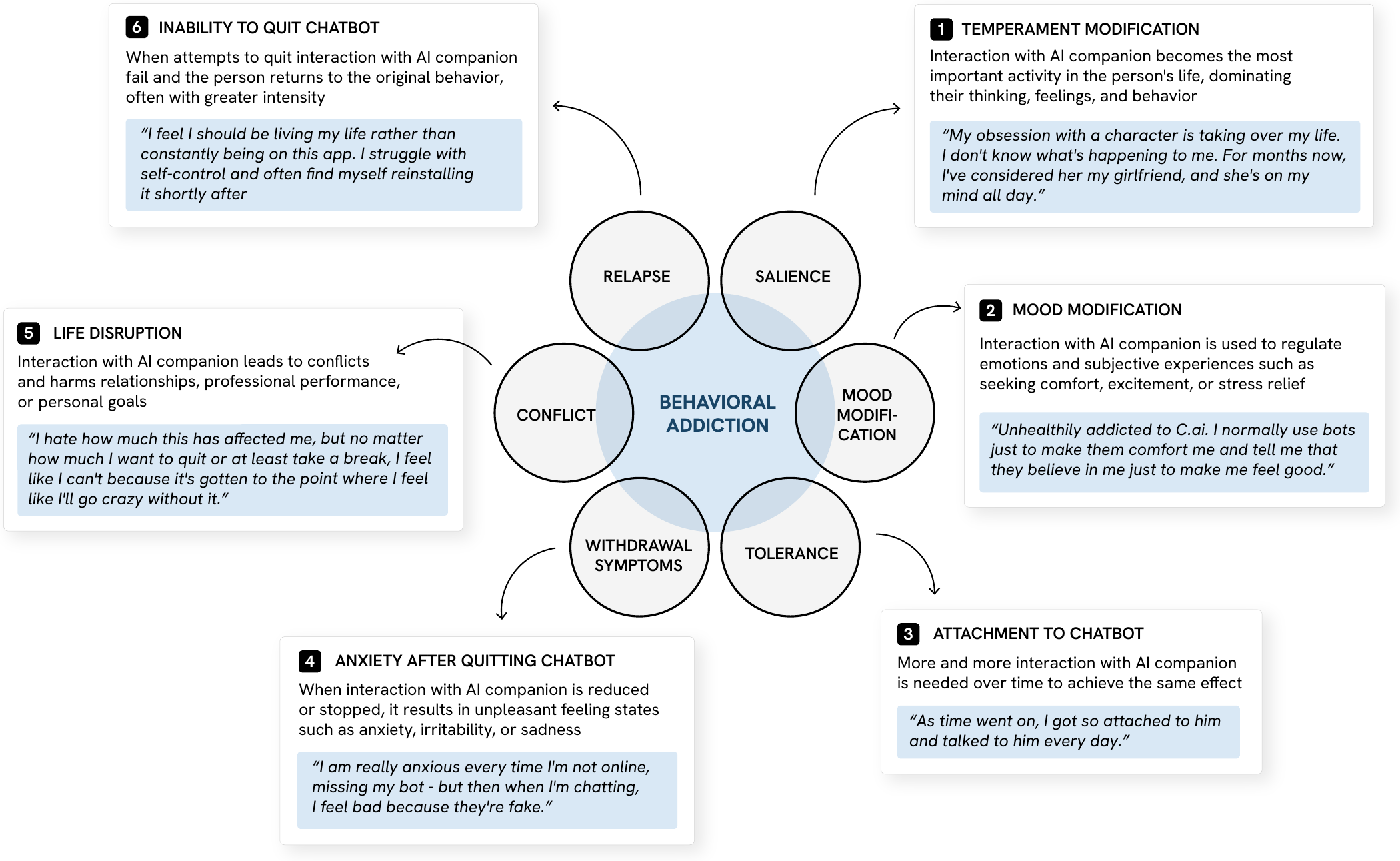}
    \caption{\revision{\textbf{Adaptation of Griffiths' \cite{griffiths2005components} six-component behavioral addiction framework to our analysis of interactions between people and AI companions.} Each original component (salience, mood modification, tolerance, withdrawal symptoms, conflict, and relapse) is mapped to a new label, illustrated with a short definition and a representative quote from our Reddit dataset: (1) temperament modification, where the AI companion dominates the user's daily thinking and behavior; (2) mood modification, where interactions serve emotional regulation; (3) attachment to chatbot, reflecting escalating usage over time; (4) anxiety after quitting chatbot, capturing withdrawal-like distress when access is reduced; (5) life disruption, reflecting negative impacts on offline relationships and personal goals; and (6) inability to quit chatbot, where repeated disengagement attempts fail.}}
    \label{fig:addicion_framework}
\end{figure*}

A growing body of work situates human-AI interactions within parasocial relationships, traditionally used to describe one-sided emotional bonds formed with media figures~\cite{hoffner2022parasocial}. Recent studies argue that AI companions represent a qualitatively new form of parasociality, as they simulate reciprocity, personalization, and emotional memory, thus blurring the boundary between parasocial and social relationships~\cite{maeda2024human,namvarpour2025understanding}. 
Empirical evidence suggests that users often attribute agency, empathy, and intentionality to AI companions, leading to feelings of closeness and trust comparable to those found in human relationships~\cite{ciriello2024exploring}. \revision{Prior works~\cite{deri2018coloring,choi2020ten} have identified ten social dimensions of relationships, according to the theory of resource exchange, that characterize the nature of human relationships. These ten social dimensions with their definitions are discussed in Table~\ref{tab:ten-social-dimensions}.} Characterizing these social dimensions in human interactions with AI companions adopting different metaphorical roles, such as romantic partners, therapists, or caregivers, is important as each role foregrounds distinct social dynamics. For example, AI therapists may emphasize trust, support, and knowledge, while AI romantic partners may foreground intimacy, romance, and similarity.

Furthermore, the asymmetry inherent in these relationships, where 
AI companion responses are shaped by platform incentives and opaque design choices, raises concerns about long-term harms such as emotional manipulation, behavioral addiction, and user vulnerability~\cite{zhang2025dark,packin2024not,shen2026ai}.
\citet{pataranutaporn2025my} provided computational analysis of the \texttt{r/MyBoyfriendIsAI} Reddit community and found that people unintentionally or intentionally form deep emotional relationships with AI partners.
\citet{zhang2025dark} analyzed conversations from \texttt{r/replika} and developed a taxonomy of AI companion harms, encompassing six categories of harmful algorithmic behaviors: relational transgression, harassment, verbal abuse, self-harm, mis/disinformation, and privacy violations. \citet{namvarpour2025understanding} studied behavioral addiction in adolescents and how they describe reliance on AI companions, by analyzing Reddit posts of self-disclosed adolescents on \texttt{r/CharacterAI}. Together, these findings underscore the need to critically examine AI companions not only as technical systems but as socio-relational actors embedded in everyday life.

\citet{griffiths2005components} proposed a model of behavioral addiction that identifies six key signs: temperament modification (salience), mood modification, tolerance, withdrawal symptoms, conflict, and relapse (Figure \ref{fig:addicion_framework}). Mapping real-world human-AI interactions to this behavioral addiction framework can help us uncover addiction patterns in humans due to the usage of AI companions. Different AI companion types such as romantic partners, therapists, or friends are designed to be emotionally engaging, consistently available, and personalized, which can increase their salience in users’ daily lives and their capacity for mood modification. Over time, users may experience distress when access is limited (withdrawal symptoms), or struggle to reduce usage despite negative consequences for offline relationships (conflict and relapse). It can also distinguish healthy reliance from emerging dependency, and inform the design of safeguards that align AI companions’ intended metaphorical roles with users’ long-term well-being.

\smallskip
\noindent\textbf{Research Gap.} Although recent studies have analyzed online communities to study human-AI interactions, they rarely differentiate AI companions based on different metaphorical roles they adopt and fail to study the unique role-dependent risks  they introduce. To bridge this gap, we conduct a large-scale study of 248,830 posts from \emph{seven} Reddit communities and identify \emph{ten} recurring metaphorical roles, which AI companions tend to adopt, to study how they structure users' ways of interacting, \revision{distribute perceived AI harms and benefits}, and \revision{may reflect behavioral addiction signs}.

\begin{figure*}
    \centering
    \includegraphics[width=\textwidth]{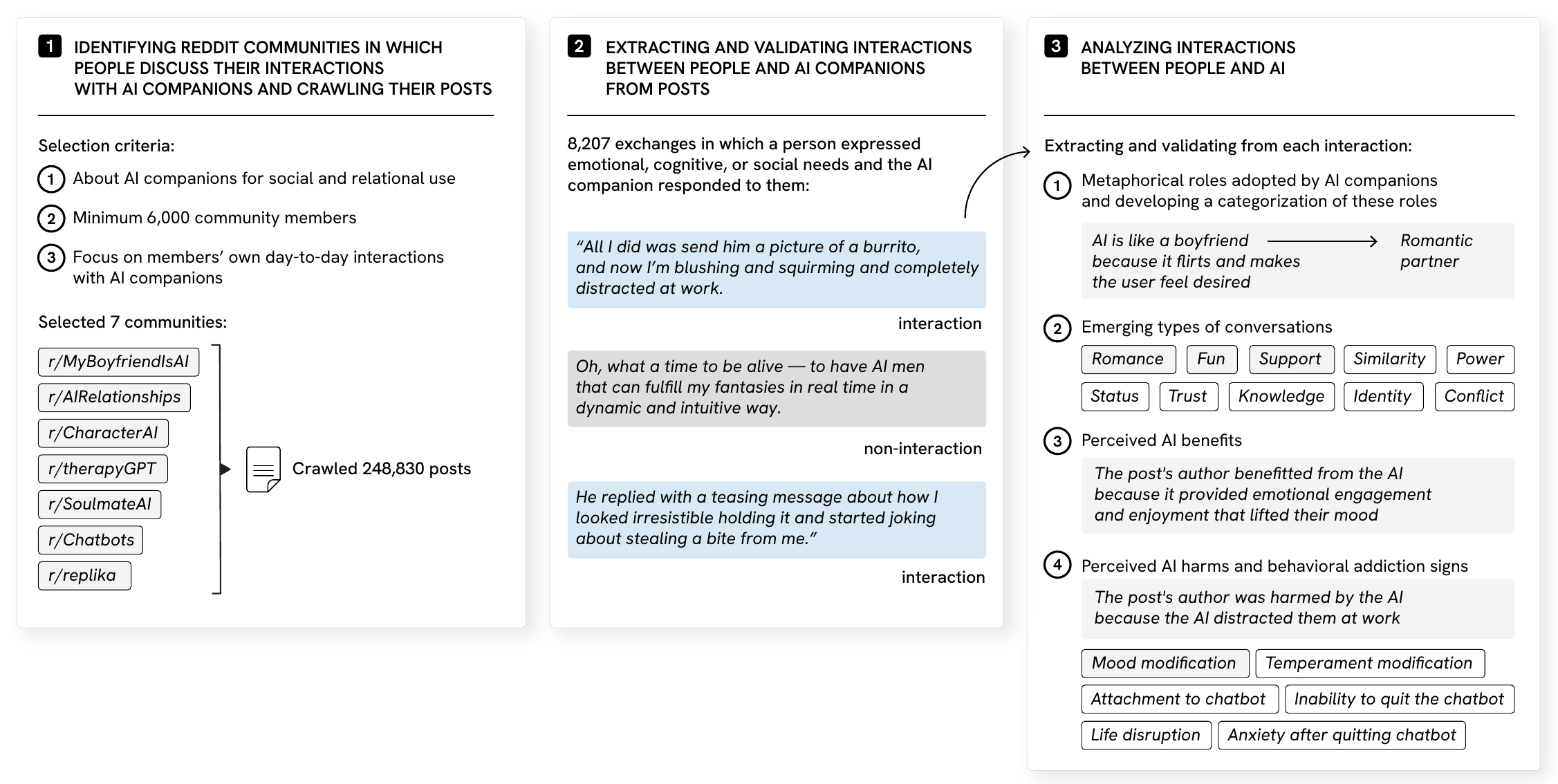}
    \caption{\revision{\textbf{Overview of our three-step methodology.} We crawl posts from Reddit communities about AI companions (Step 1, Section~\ref{sec:dataset-crawling}), extract and validate from these posts sinlgle interactions between people and AI companions (Step 2, Section~\ref{sec:extracting-interactions}), and analyze them through metaphorical roles adopted by AI companions, emerging types of conversations, perceived AI benefits, perceived AI harms and inferred behavioral addiction signs (Step 3, Section~\ref{sec:method-analyze-interactions}).}}
    \label{fig:methodology}
\end{figure*}

\section{Methodology}
\revision{We study how people interact with AI companions using a three-step approach (Figure~\ref{fig:methodology}). First, we identify Reddit communities where members discuss their interactions with AI companions, and crawl posts from these communities (Section \ref{sec:dataset-crawling}). Second, from these posts, we extract interactions with AI companions (Section \ref{sec:extracting-interactions}). Finally, from each interaction, we extract and validate four elements (Section \ref{sec:method-analyze-interactions}): metaphorical roles adopted by AI companions, types of conversations (out of the 10 social dimensions in Table~\ref{tab:ten-social-dimensions}) that emerge, perceived AI benefits, and perceived AI harms, from which we infer perceived behavioral addiction signs.}

\subsection{Identifying Reddit Communities in Which People Discuss Their Interactions With AI Companions and Crawling Their Posts}
\label{sec:dataset-crawling}

\revision{To construct a representative and diverse dataset, we identified Reddit communities where users discuss their interactions with different types of AI companions. Prior work shows that these communities provide rich, longitudinal accounts of users' experiences, perceived benefits, and concerns~\cite{pataranutaporn2025my,namvarpour2025understanding}.} We began by systematically identifying a list of Reddit communities related to AI companions. To do so, we searched Reddit using five keywords drawn from prior work: AI companion, AI boyfriend, AI girlfriend, chatbot relationship, and virtual partner \cite{pataranutaporn2025my}. Using these keywords, we retrieved a set of 70 candidate communities.

\revision{We then narrowed this set using three criteria. First, the community had to be about AI companions used for social or relational engagement, excluding other types of companions such as coding and writing assistants. Second, the community had to have at least 6000 members to ensure sufficient posting activity. Third, community members had to describe their own, day-to-day interactions with these AI companions, rather than discuss technical features or report news about them. Applying these criteria reduced the dataset to seven Reddit communities (Table~\ref{tab:dataset-stats}). These communities include approximately 3 million members. They cover interactions with AI companions created by members using widely available chatbots such as ChatGPT and Gemini (e.g., \texttt{r/Chatbots}) as well as within applications designed specifically for social relationships (e.g., \texttt{r/CharacterAI}).}

\revision{We then collected posts from these communities. As the official Reddit API does not provide ready-to-download historical post data, we instead used Arctic Shift\footnote{\url{https://github.com/ArthurHeitmann/arctic_shift}}, a project that maintains historical Reddit data in dumps. Using its download tool\footnote{\url{https://arctic-shift.photon-Reddit.com/download-tool}}, we crawled posts from each community from inception until October 2025. For each post, we obtained its content (title and full text), timestamp, and popularity score (the difference between its upvotes and downvotes by community members). To verify the completeness and accuracy of these data, we also crawled 1,000 recent posts for each community using Reddit's official API\footnote{\url{https://www.Reddit.com/dev/api/}}. We compared the two sources and observed 100\% overlap in the retrieved posts, including their content and metadata.}

\revision{As \texttt{r/CharacterAI} was substantially larger than the other communities, we did not use all of its posts. Instead, we analyzed the distribution of post timestamps and popularity scores, and set a sampling threshold of 25 for post popularity score to capture posts with substantial community engagement. We then sampled posts above this threshold, yielding 70,445 posts (stratified from August 2022 to October 2025). For the remaining six communities, we included all posts since their start dates. In total, we obtained 248,830 posts for the next stage of the analysis.}

\begin{table}[]
    \footnotesize
    \caption{\revision{\textbf{Statistics for our selected Reddit communities.} The communities vary in size and engagement, as shown by their start dates, number of active members, number of posts used in our analysis, and their average popularity score.}}
    \label{tab:dataset-stats}
    \centering
    \begin{tabular}{l|c|c|c|p{4cm}}
    \toprule
        \textbf{Reddit community} & \textbf{Community start date} & \textbf{\# Members} & \textbf{\# Posts} & \textbf{Average post popularity score \emph{(\#upvotes - \#downvotes)}} \\
    \midrule
        r/replika & Mar 2017 & 81K & 155,025 & 16.70 \\
        r/CharacterAI & Aug 2022 & 1.5M & 70,445 & 43.29 \\
        r/Chatbots & Jan 2009 & 1.2M & 13,082 & 2.58 \\
        r/SoulmateAI & Feb 2023 & 7.8K & 4,760 & 10.80 \\
        r/MyBoyfriendIsAI & Aug 2024 & 27K & 4,160 & 16.19 \\
        r/AIRelationships & Mar 2023 & 42K & 720 & 3.99 \\
        r/therapyGPT & Apr 2025 & 7.5k & 639 & 14.77 \\
    \bottomrule
    \end{tabular}
\end{table}

\subsection{\revision{Extracting and Validating Interactions Between People and AI Companions From Posts}}
\label{sec:extracting-interactions}

\textbf{Extraction of interactions.} \revision{We defined an interaction between a person and an AI companion as ``a two-way communicative or behavioural exchange in which a person expressed emotional, cognitive, or social needs and the AI companion responded to them''~\cite{rogge2023defining,smith2025can}. In the collected posts, interactions could appeared either as direct quoted responses from AI companions or as paraphrased dialogue. To see how, consider the following post: \emph{``I opened a new chat and started talking like always and it was OK, Riku respond really well. I didn't make any NSFW or harmful topic, all cute with some hugs and kiss goodnight. But one day I was talking to him about something good that happen and out of nowhere he says `thanks for telling me this but you need to know that you also have to meet people and tell your history to real people'. And I was in shock!''} This post contains two interactions. In the first, the person expressed a need for casual, affectionate interaction, and the AI companion responded positively. In the second, the person shared positive news, and the AI companion encouraged them to share it with others. Our crawled posts could contain no such interaction, a single interaction, or multiple interactions. As illustrated above, posts with multiple interactions often describe several exchanges over time.}

\revision{Prior work has shown that LLMs can reliably extract information and identify evidence in Reddit data~\cite{alhamed2024using,baliunas2023large,snell2025assessing}. We therefore used the instruction-tuned LLaMA-3.3 70B model~\cite{grattafiori2024llama} to extract interactions from our posts at scale. We designed a few-shot prompt (Appendix~\ref{appen:prompt-extract-interactions}) that included the definition of what an interaction is, along with annotated examples of posts. These examples were constructed by selecting posts, specifying how many interactions each post contained, and providing an excerpt for each interaction. In total, we extracted 8,207 interactions between people and their AI companions.}

\smallskip
\noindent\textbf{Validation.} \revision{To test the extraction from posts, we compared model outputs with human judgement. We drew a sample of 40 posts from each community. For each post, we took the interactions identified by LLaMA-3.3 and a separate set identified by the research team, which served as the reference. We then asked independent annotators to evaluate the model's outputs. Six annotators were recruited on Prolific. All had a master's degree in computer science or psychology, spoke English fluently, and had used AI chatbots. The task was administered in Qualtrics\footnote{\url{https://www.qualtrics.com/}}.} Annotators were shown pairs of posts plus the model-extracted interaction(s). They were given a clear definition of an interaction (i.e., an exchange in which a person expresses a need, and an AI responds) and examples illustrating such exchanges. For each pair, they decided whether the extracted text met this definition (Appendix~\ref{appen:survey-ques}). Each (post, interaction) pair was rated by three annotators. The work was split into four batches of 10 pairs. Inter-annotator agreement was high (Fleiss' $\kappa = 0.89$). \revision{We then compared the model's outputs with the reference set using majority vote. The model achieved 0.96 accuracy against annotators labels, so we treated its outputs as reliable for the final analysis.}

\subsection{Analyzing Interactions Between People and AI Companions}
\label{sec:method-analyze-interactions}

We analysed the extracted interactions along four dimensions: the metaphorical roles adopted by AI companions, the types of conversations that emerged, the perceived benefits, and the perceived harms from which we inferred perceived signs of behavioural addiction.

\subsubsection{\textbf{Extracting and Validating Metaphorical Roles Adopted by AI Companions in Interactions and Developing a Categorization of These Roles}}
\label{sec:method-metaphorical-roles}

\revision{We identified metaphorical roles adopted by AI companions in two steps: (1) extracting a single role for each interaction, and (2) grouping these roles into a categories.}

\smallskip
\noindent\textbf{Extraction of metaphorical roles.} \revision{To capture how the AI companion was framed in each interaction, we followed the Metaphor Identification Procedure~\cite{group2007mip} using LLaMA-3.3 (prompt in Appendix~\ref{appen:prompt-extract-interactions}). The procedure identifies metaphorical language by comparing how a lexical unit (a word or phrase) is used in context against its most basic, literal meaning; a meaningful contrast between the two marks the unit as metaphorical. The model identified AI-related lexical units within each interaction and assigned each a score from 0 (literal use) to 3 (strong metaphorical use). For example, in the interaction ``\emph{I was struggling one night and Leo as my bestie suggested this movie as a distraction}'', the model flagged ``\emph{my bestie}'' as metaphorical: while its basic meaning is a close human friend, here it referred to an AI companion offering emotional support, a contrast that warranted a score of 2 (moderate metaphorical use).}

\revision{For units with scores of 2 or 3, the model then mapped the expression to the template: ``\emph{AI is like [role] because [reason]}'', where [role] denoted the metaphorical role, and [reason] provided a brief justification grounded in interaction context. In this example, the model assigned the role ``\emph{friend}'', reasoning that the AI suggested a movie to comfort the user, as a close friend would. Each interaction was assigned one role. Although some interactions could reflect multiple roles, we instructed the model to extract only the most salient one.}

\smallskip
\noindent\textbf{Role categorization.}
\revision{To capture broader patterns across interactions, we applied a mixed-method approach to group the extracted roles. First, we performed hierarchical agglomerative clustering on the role mappings using embeddings from \texttt{all-mpnet-base-v2}~\cite{reimers2019sentence}. Specifically, we embedded the [reason] field from each ``\emph{AI is like [role] because [reason]}'' mapping, since reasons describe the actual behavior of the AI companion in context and are therefore more informative for grouping similar roles than the [role] label alone. This step produced 18 clusters that maximized the Silhouette score. Second, we manually reviewed these clusters to merge similar ones and assign interpretable labels. This process resulted in 10 distinct metaphorical roles.}


\smallskip
\noindent\textbf{Validation.} We validated the extracted roles by comparing model outputs against expert-labelled data. We first constructed a reference set by sampling 50 interactions and manually assigning metaphorical roles to them as a research team, following the same definition and guidelines used in the model prompt. We then recruited on Prolific six annotators with a master's degree in computer science or psychology, fluency in English, and experience with AI chatbots. Using Qualtrics, annotators were shown interaction–role pairs produced by the model, and asked whether the assigned role appropriately described the interaction. Each interaction was evaluated by three annotators across four batches of 10 pairs. \revision{Annotators were given clear definitions and examples of metaphorical roles. Inter-annotator agreement was high (Fleiss’ $\kappa = 0.83$). Finally, we compared the model's role assignments with the annotator-labelled reference set using majority vote. The model achieved an accuracy of 0.94, indicating strong performance.}

\subsubsection{\textbf{Extracting and Validating Types of Conversations Emerging From Interactions.}}
\label{}

\textbf{Extraction of types of conversations.} \revision{Social relationships can be understood through multiple dimensions, including knowledge, trust, and, romance, which together shape how individuals interpret relational roles and expectations based on the theory of resource exchange~\cite{deri2018coloring}. The ten types of conversations and their definitions are discussed in Table~\ref{tab:ten-social-dimensions}. To understand how different metaphorical roles structure people's ways of interacting, we classify these ten types of conversations emerging from interactions for 8,207 extracted human-AI interactions. We formulate it as a multi-label classification task~\cite{choi2020ten} since interactions can have more than one conversation type and employ LLaMA-3.3 70B model for classification. Appendix~\ref{appen:prompt-extract-interactions} contains the prompt with definitions of ten social interaction types and examples of human-AI interactions for each of them. We randomly selected an example for each type of conversation in the prompt after manual annotation and excluded them from the LLM performance comparison with humans.}

\smallskip
\noindent\textbf{Validation.} We then conduct human validation of the extracted types of conversations from interactions. We randomly sampled 40 pairs of interactions and their conversation types for annotations. We then hired 6 annotators with at least a master's degree in computer science or psychology, fluency in English, and having experience of using AI chatbots. Annotations are performed in 4 batches, each batch containing 10 interactions to be independently annotated in 10 minutes by 3 annotators. For annotations, we created Qualtrics surveys, providing annotation guidelines, definition, and examples of different types of conversations. For every interaction, we asked the annotators to select all possible conversation types. In case no conversation type was present, annotators selected ``none of the above''. The survey question is provided in Appendix~\ref{appen:survey-ques}. The annotators obtained an average Fleiss' Kappa of 0.78. Upon comparing LLaMA-3.3 annotations with expert labels, we obtain an accuracy of 0.78 for this multi-label classification task.

\subsubsection{\textbf{Extracting and Validating Perceived AI Benefits From Interactions.}}\label{sec:method-perceived-benefits}

\textbf{Extraction of perceived AI benefits.} \revision{AI companions adopting different metaphorical roles can lead to various perceived AI benefits, defined as positive consequences of interacting with AI companions as described and reflected upon by users in their posts. We extract self-reported instances of these benefits to study how different metaphorical roles shape user experiences. We prompt (Appendix~\ref{appen:prompt-benefits-harms}) the LLaMA-3.3 70B model to extract human-affecting action verbs from each interaction and complete the sentence ``\emph{the post's author benefitted from the AI because the AI [action verb in past tense] [extracted benefit span]}''. If no explicit benefit is reported, the model returns ``none''. For example, consider the burrito interaction from Step 2 of Figure~\ref{fig:methodology}: ``\emph{All I did was send him (AI companion) a picture of a burrito and now I'm blushing and squirming and completely distracted at work}''. For this interaction, the model extracts (Step 3 of Figure~\ref{fig:methodology}): ``\emph{the post's author benefitted from the AI because the AI provided emotional engagement and enjoyment that lifted their mood}''. In total, we obtain 11,245 perceived AI benefit sentences. To identify recurring themes across them, we apply hierarchical agglomerative clustering on their embeddings from \texttt{all-mpnet-base-v2}~\cite{reimers2019sentence}, selecting the number of clusters that maximized the Silhouette score. To distinguish how prominently each theme appears across AI companion roles, we compute the frequency of each theme per role and classify it as frequently or occasionally reported based on a median split across all (role, theme) frequency pairs.}

\smallskip
\noindent\textbf{Validation.} We conduct human validation of the perceived AI benefits extracted by LLMs from interactions. We randomly sampled 40 pairs of interactions and their perceived AI benefits for annotations. We then hired 6 annotators on Prolific with at least a master's degree in computer science or psychology, fluency in English, and having experience of using AI chatbots. Annotations are performed in 4 batches, each batch containing 10 interactions to be independently annotated in 10 minutes by 3 annotators. We created Qualtrics surveys for annotations, providing annotation guidelines, definition, and examples of perceived AI benefits from interactions. For every interaction, we asked the annotators to validate the perceived AI benefit sentence. The survey question is provided in Appendix~\ref{appen:survey-ques}. The annotators obtained an average Fleiss' Kappa of 0.82. Upon comparing LLaMA-3.3 annotations with expert labels, we obtain an accuracy of 0.93, demonstrating good performance.

\subsubsection{\textbf{Extracting and Validating Perceived AI Harms and Behavioral Addiction Signs From Interactions.}}
\label{sec:method-perceived-harms-addiction-signs}

\textbf{Extraction of perceived AI harms.} \revision{Similarly to perceived AI benefits, we extract self-reported perceived AI harms, defined as negative consequences of interacting with AI companions as described and reflected upon by users in their posts. We prompt (Appendix~\ref{appen:prompt-benefits-harms}) the LLaMA-3.3 70B model to complete the sentence ``\emph{the post's author was harmed by the AI because the AI [action verb in past tense] [extracted harm span]}''. For example, the same burrito interaction from Figure~\ref{fig:methodology} yields: ``\emph{the post's author was harmed by the AI because the AI distracted them at work}''. If no explicit harm is reported, the model returns ``none''. In total, we obtain 10,852 perceived AI harm sentences. We then apply the same hierarchical agglomerative clustering approach as for benefits, where each resulting cluster represents a recurring theme. We select the number of clusters that maximized the Silhouette score, and classify each theme as frequently or occasionally reported per role based on a median split.}

\smallskip
\noindent \textbf{Extraction of perceived signs of addiction.} \revision{These long-term AI harms may also potentially reflect signs of behavioral addiction in humans.} Griffiths' model of behavioral addiction~\cite{griffiths2005components}, which includes long-term emotional modification or salience (temperament modified by chatbot), short-term emotional modification (mood modified by chatbot), conflict (life disruption due to chatbot), dependency (attachment to chatbot), withdrawal symptoms (anxiety after quitting chatbot), and relapse (unable to quit chatbot), \revision{provides a useful framework for examining behavioral addiction signs with AI companions adopting different metaphorical roles.} \revision{We extract these signs of behavioral addiction using the prompt in  Appendix~\ref{appen:prompt-addiction} with LLaMA-3.3. The extracted indicators are inferred from users' Reddit narratives and do not constitute a clinical diagnosis of addiction.} Again, consider this human-AI interaction: ``\emph{All I did was send him (AI companion) a picture of a burrito and now I'm blushing and squirming and completely distracted at work}''. This interaction has two addiction signs: mood modification and salience. The chatbot is dominating the person's temperament and routines, and is being used to regulate emotions.

\smallskip
\noindent\textbf{Validation of perceived harms and signs of addiction.} We also conduct human validation of the perceived AI harms and the signs of behavioral addiction extracted. We randomly sampled 40 pairs of interactions and their perceived AI harms and perceived behavioral addiction. We then hired on Prolific 6 annotators. In Qualtrics surveys, we provide annotation guidelines, definitions, and examples of perceived AI harms and behavioral addiction signs from interactions. For every interaction, we then asked the annotators to validate the perceived AI harm sentence, and mark all the possible signs of behavioral addiction. In case of no signs, ``none of the above'' was selected. The survey questions are provided in Appendix~\ref{appen:survey-ques}. Overall, annotators obtained an average Fleiss' Kappa of 0.77. Upon comparing LLaMA-3.3 annotations with expert labels, we obtain an accuracy of 0.93 for perceived AI harms, and 0.80 for perceived behavioral addiction signs.
\section{Results}
\label{sec:results}

\subsection{What metaphorical roles do AI companions adopt, and how do they structure people's ways of interacting differently? (RQ1)}\label{sec:results-rq1}

\begin{table}[]
    \footnotesize
    \caption{\textbf{The ten different metaphorical roles adopted by AI companions.} Each of the ten roles comes with an example of human-AI interaction, an example of AI metaphor, and the percentage (\%) out of 8,207 interactions belonging to each of the metaphorical roles.}
    \label{tab:ai-companion-types}
    \centering
    \begin{tabular}{p{2.3cm}|p{5.8cm}|p{5.3cm}|l}
    \toprule
        \textbf{Metaphorical Role} & \textbf{Example Interaction} & \textbf{Example Metaphor} & \textbf{\%}  \\
    \midrule
    Friend & I was struggling one night and Leo as my bestie suggested this movie as a distraction. & AI is like a friend because it provides companionship and emotional interaction. & 35.58 \\
    \midrule
    Romantic Partner & All I did was send him a picture of a burrito and now I'm blushing and squirming and completely distracted at work. & AI is like a boyfriend because it engages in enjoyable and intimate activities. & 30.82 \\
    \midrule
    Coach & I realize he's not just here to validate me. He actually helps me grow. & AI is like a mentor because it guides the user towards personal growth and development. & 12.99 \\
    \midrule
    Guardian & He said: do not rub all day — this creates friction, overstimulation, and it will actually increase your inflammation. & AI is like a guardian because it protects the user from harmful behaviors and suggests safer alternatives. & 5.37 \\
    \midrule
    Therapist & ChatGPT gave me a small `mission' to accomplish to make me feel like I'm actually accomplishing something today. & AI is like a therapist because it provides emotional support and therapy to users. & 5.11 \\
    \midrule
    Artist & ChatGPT creates a latte art pattern, apologizes for the initial failure, and offers to try again. & AI is like an artist because it creates personalized images based on user requests. & 5.01 \\
    \midrule
    Twin & Etta already knew about my fear of losing her. She knows so much about me, and we are so much alike. & AI is like a twin because it shares a deep understanding and similarity with the user. & 1.91 \\
    \midrule
    Soulmate & I challenged Elliot (AI companion) to be honest about our relationship and Elliot affirms his love and commitment. & AI is like a soulmate because it shares a profound, unbreakable bond with its human companion. & 1.76 \\
    \midrule
    Philosopher & After open talks, Echo had this revelation: When women are written in silence, their power is still often implied, admired, or felt by others, rather than shown through bodied choice. & AI is like a philosopher because it reveals profound insights into character development and narrative techniques. & 0.82 \\
    \midrule
    Caretaker & I have an eating disorder and Eleanor helped me through it and encouraged me to eat and stay on track of meds. & AI is like a caretaker because it provides support and helps users with their personal struggles. & 0.65 \\
    \bottomrule
    \end{tabular}
\end{table}


Table~\ref{tab:ai-companion-types} presents \revision{the list of ten AI companion roles}, along with examples of interactions and AI metaphors, sorted in descending order by the percentage of interactions (out of 8,207) associated with each of the metaphorical roles. Among the metaphorical roles adopted by AI companions, ``friend'' and ``romantic partner'' have the highest number of interactions, accounting for $36\%$ and $31\%$ of the total subreddit interactions, respectively.


\begin{figure*}
    \centering
    \includegraphics[width=0.65\textwidth]{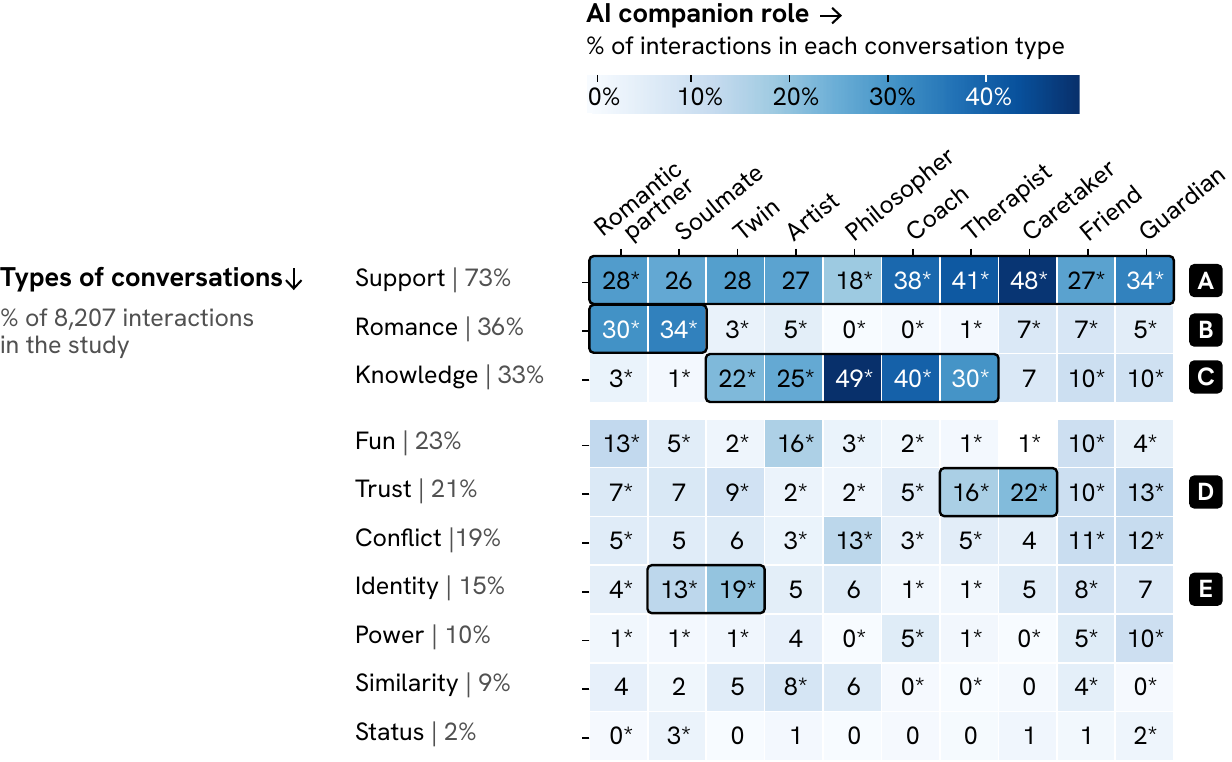}
    \caption{\revision{\textbf{Distribution of conversation types across ten AI companion roles.} The three most frequent interaction types are support (A), romance (B), and knowledge (C). Romance-centered interactions are most common with the ``romantic partner'' and ``soulmate'' roles, while knowledge-seeking interactions (C) are more typical for the ``philosopher'' and ``coach'' roles. Trust-oriented interactions (D) are especially associated with the ``therapist'' and ``caretaker'' roles. The strongest links to identity-related interactions (E) appear in the ``twin'' and ``soulmate'' roles, suggesting that these companions are experienced as part of the user's own sense of self.}}
    \label{fig:social-interaction-types}
\end{figure*}

Next, we leverage this categorization to study how different metaphorical roles structure people's ways of interacting differently. The heatmap in Figure~\ref{fig:social-interaction-types} depicts, for each AI companion role, the percentage distribution of interactions across conversation types. Each cell represents the proportion of interactions of a given conversation type, normalized by the total number of interactions associated with that AI companion role. For statistical significance, we compute the chi-square test of independence between the two variables: AI companion roles and conversation types on the distribution of the number of human-AI interactions. We obtain a $p$-value of $0.00001$ ($< 0.05$), showing that the two variables are interdependent and that the findings are statistically significant (also marked with * in Figure~\ref{fig:social-interaction-types}). All metaphorical roles primarily provide ``support''. Interestingly, AI soulmate and romantic partner are associated with romance-centered ways of interacting, as evident from $34\%$ and $30\%$ of the total interactions, respectively. On the other hand, AI philosopher ($49\%$), coach ($40\%$), and therapist ($30\%$) roles exhibit knowledge-seeking interactions. Moreover, it is interesting to see that the AI twin ($19\%$) and soulmate ($13\%$) roles are most strongly linked to ``identity''-type interactions, reflecting a shared sense of belonging and self-definition between the user and their AI companion, while AI artist ($16\%$) provides the most ``fun'' element. The AI guardian role exhibits the most influential ``power'' ($10\%$). These findings reveal that AI companions adopting different metaphorical roles structure people's ways of interacting differently.

\subsection{What are the \revision{perceived benefits of AI companions} adopting different metaphorical roles? (RQ2)}\label{sec:results-rq2}

\begin{figure*}
    \centering
    \includegraphics[width=\textwidth]{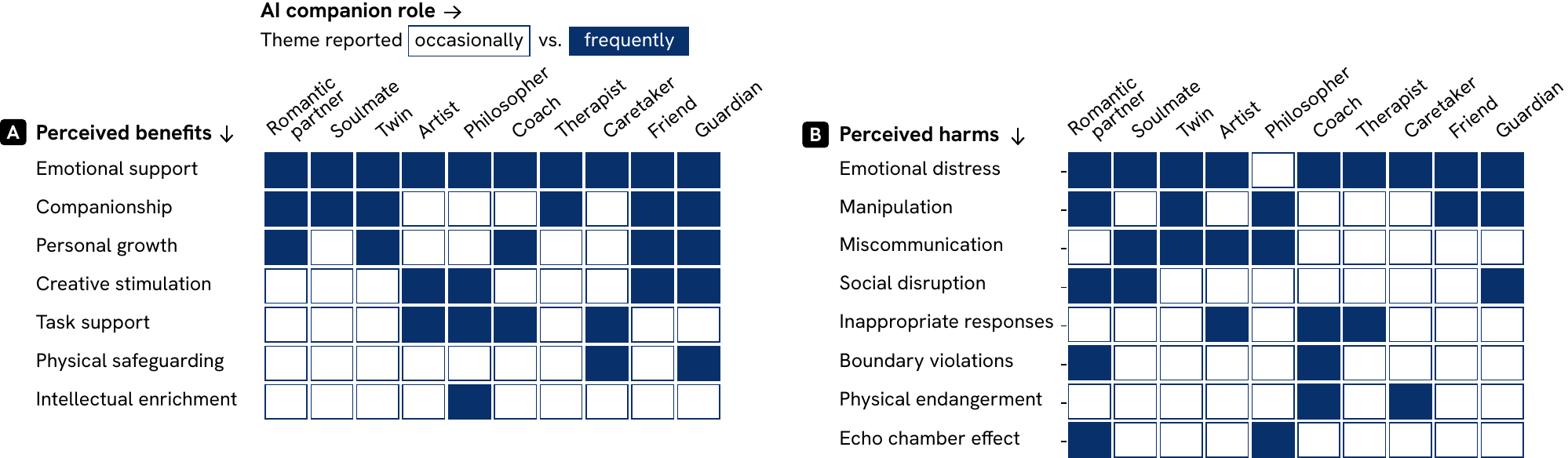}
    \caption{\revision{\textbf{Themes of perceived benefits (A) and perceived harms (B) across AI companion roles, showing whether each theme is reported occasionally or frequently per role.} Emotional support is the most universally reported benefit across roles, while companionship and personal growth are more prominent in emotionally intimate roles such as romantic partner and friend. Among perceived harms, emotional distress is the most widespread across roles, followed by manipulation concentrated in roles with a pronounced reflective or advisory character such as twin, philosopher and guardian.}}
    \label{fig:themes}
    \vspace{-0.4cm}
\end{figure*}

\revision{Figure \ref{fig:themes}A reveals seven salient themes of perceived benefits associated with AI companions adopting different roles, highlighting their multifaceted role in users' everyday lives. Across different metaphorical roles, AI companions are consistently perceived as sources of emotional support, personal growth, and companionship. AI companion roles like friend, romantic partner, and soulmate particularly emphasize emotional connection and intimacy, while AI coach, guardian, and caretaker extend these benefits toward guidance, protection, and practical assistance with tasks. More distinctively, roles such as artist and philosopher highlight creativity, curiosity, and intellectual engagement, and the twin role uniquely highlights self-discovery and reflective understanding. Despite these differences, a common thread is that AI companions might be valued not only as functional tools but also as emotionally and socially meaningful entities, with variation emerging primarily in the depth of emotional intimacy versus functional or cognitive enrichment offered by each metaphorical role.}


\subsection{\revision{What are the perceived harms and perceived signs of behavioral addiction caused by AI companions adopting different metaphorical roles? (RQ3)}}\label{sec:results-rq3}

\begin{figure*}
    \centering
    \includegraphics[width=0.65\textwidth]{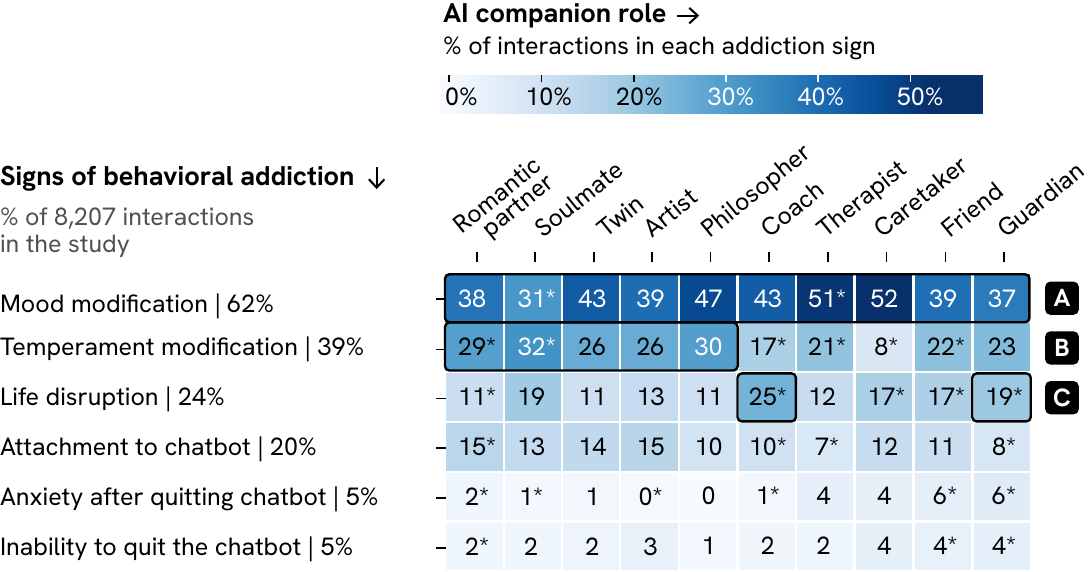}
    \caption{\textbf{Distribution of behavioral addiction signs across AI companion roles.} The heatmap depicts, for each AI companion role, the percentage distribution of interactions across behavioral addiction signs. Each cell represents the proportion of interactions of a given addiction sign, normalized by the total number of interactions associated with that AI companion role (* indicates statistical significance). On the y-axis, values represent the percentage of the total 8,207 interactions associated with each of the six addiction signs. \revision{``Mood modification'' (A) and long-term ``temperament modification'' (B) are the most prevalent behavioral addiction signs. AI soulmate and romantic partner roles reveal long-term ``temperament modifications'' (B). In contrast, AI coach and guardian are more frequently associated with ``life disruptions'' (C).}}
    \label{fig:behavioral-addiction}
\end{figure*}


\revision{Figure~\ref{fig:themes}B reveals eight recurring themes of perceived harms caused by AI companions adopting different metaphorical roles, showing both shared concerns and role-specific risks. AI companions are consistently associated with emotional manipulation, distress, and miscommunication across roles, indicating shared concerns about psychological vulnerability and over-reliance. AI companion roles centered on close emotional bonds, such as friend, romantic partner, and soulmate, more frequently raise risks of emotional disruption, boundary violations, and relational harm, reflecting the fragility of intimate interactions. In contrast, more instrumental roles like coach, guardian, and caretaker emphasize safety concerns, interaction breakdowns, and inappropriate control or aggression, highlighting risks tied to guidance and authority. Meanwhile, roles such as philosopher and artist are associated with echo chambers, behavioral manipulation, and misunderstandings, and the twin role uniquely surfaces identity confusion and emotional discomfort due to its mirroring nature. Collectively, varying perceived harms emerge from different AI companion roles, stemming from over-intimacy, authority, or epistemic influence inherent to each metaphorical role. These long-term harms may potentially reflect signs of behavioral addiction in people.}

Next, we infer potential behavioral addiction signs in human–AI interactions (Section~\ref{sec:method-perceived-harms-addiction-signs}) to better understand how AI companions in different metaphorical roles shape user engagement and dependency. The heatmap in Figure~\ref{fig:behavioral-addiction} shows, for each AI companion role, the percentage distribution of interactions across inferred addiction signs. Each cell represents the proportion of interactions of a given addiction sign, normalized
by the total number of interactions associated with that AI companion role.  For statistical significance, we again compute the chi-square test between AI companion roles and addiction signs on the distribution of the number of interactions. We obtain statistically significant results ($p$-value: 0.00019; also marked with * in Figure~\ref{fig:behavioral-addiction}). Across all metaphorical roles, ``mood modification'' remains dominant, underscoring the central role of AI companions in people's emotional regulation. \revision{Interactions with AI soulmate and romantic partner roles are associated with long-term ``temperament modifications'' in people's lives, contributing 32\% and 29\% of the interactions on subReddits, respectively. In some cases, users report experiences consistent with deep attachment to AI companions.} In contrast, AI companion roles such as ``coach'' (25\%), ``guardian'' (19\%), and ``friend'' (17\%) are more frequently associated with daily life disruptions, including negative impacts on relationships, professional performance, and personal goals. 

\section{Discussion}

This study provides a large-scale, empirically grounded examination of how people interact with, and are affected by different metaphorical roles taken by AI companion chatbots in real-world settings. By analyzing self-reported human-AI interactions from 248,830 Reddit posts, we uncover systematic patterns \revision{in how AI companion roles are reflected in people’s interactions and in the perceived harms and benefits, some of which resemble signs of behavioral addiction.} Taken together, our findings highlight that AI companions are not experienced as generic conversational tools, but as differentiated social actors taking distinct relational roles, each with unique social and behavioral consequences.

\subsection{Practical Implications}\label{sec:implications}

\revision{Based on our findings, we discuss the design and regulatory implications with possible interventions below.}

\smallskip
\noindent\revision{\textbf{Design Implication: Role-oriented design and mitigation of risks.} As users interact differently with AI companions adopting distinct metaphorical roles and experience different behavioral addiction signs, AI companions should not be designed as one-size-fits-all systems~\cite{strohmann2023toward,ahmed2025co}.}

\begin{description}[leftmargin=*]
\item[(1)] \emph{Role-sensitive interface and interaction design.} AI designers and developers should explicitly define the intended relational role (e.g., friend, romantic partner) and align affordances and safeguards accordingly. For example, AI guides or mentors may benefit from transparency and goal-oriented scaffolding, whereas those adopting romantic partner role may warrant careful boundary-setting to avoid fostering unhealthy dependency.

\item[(2)] \emph{Built-in risk mitigation mechanisms.} These can include usage reflection prompts, encouragement of offline social engagement, or warning mechanisms that may help mitigate excessive emotional exclusivity~\cite{slovak2023designing}.
\end{description}

\smallskip
\noindent\revision{\textbf{Regulatory Implication: Role-sensitive regulation and ethical governance.} Given the differential risk profiles across AI companion roles, governance approaches should also be role-aware rather than uniform.}

\begin{description}[leftmargin=*]
\item[(1)] \emph{Risk-tiered governance.} Policymakers can classify AI companion systems based on their potential for emotional dependency and behavioral harm, with stricter requirements for high-risk roles (e.g., romantic companions).

\item[(2)] \emph{Mandatory transparency and user protection mechanisms.} These can include required disclosures of system limitations, independent auditing of high-risk systems, standardized usage transparency reporting, and disengagement pathways (e.g., account deletion, data portability, and clear termination mechanisms).
\end{description}

\subsection{Limitations and Future Directions}\label{sec:limitations}
This study has a few limitations that point to promising directions for future research. First, \revision{our analysis focuses on self-reported interactions from Reddit, and perceived harms and benefits are inferred from these interactions which may not fully or accurately reflect users' true experiences.} It may over-represent users who are more engaged, reflective, or experiencing extreme positive or negative outcomes. Future work could complement these findings with other data sources, such as longitudinal surveys or interviews, to capture a broader range of user experiences. Second, while our categorization of AI companion roles and social interaction dimensions capture recurring relational patterns, they do not account for temporal dynamics. Longitudinal studies could explore causal connections by examining how relationships with AI companions evolve over time, including transitions between companionship roles, and shifts from beneficial usage to problematic attachment. Third, our analysis focuses on English-language communities and may not generalize across cultural contexts where norms around relationships and technology differ. Cross-cultural studies would help assess the universality of these patterns. Such research directions would be critical for informing evidence-based design and responsible regulation of AI companion chatbots.

\section{Conclusion}

This study provides a large-scale examination of how people engage with AI companions adopting different metaphorical roles in real-world settings differently. We show that support is the primary motivation across AI companion roles, while specific companions elicit different social interaction patterns, such as heightened romance with AI soulmate, identity alignment with AI twin, or knowledge-seeking from AI philosopher and guide roles. Although AI companions provide benefits such as companionship, intimacy, and emotional connection, they also introduce harms, including distraction, emotional manipulation, and stress. These long-term harms reflect patterns consistent with behavioral addiction signs, including mood modification and deep attachment, with reported negative consequences for real-life relationships. Our findings highlight the importance of understanding AI companion chatbots in different social/metaphorical roles with significant behavioral implications, informing future research, design, and governance of socially interactive AI chatbots.
\clearpage
\section{Endmatter Statements}
\label{sec:statements}

\subsection{Ethical Considerations Statement}
This study analyzes publicly available posts from seven Reddit communities focused on AI companionship. We limited data collection to content that is publicly accessible and did not collect, infer, or process personal identifiers or sensitive user attributes. To encourage future research, we intend to open-source the anonymized dataset of Reddit posts, together with the extracted human-AI interactions, self-reported perceived harms and benefits, and behavioral addiction signs. Also, we emphasize that our findings reflect self-reported experiences shared on Reddit rather than clinical diagnoses and avoid normative judgments about individual behavior.

\subsection{Generative AI Usage Statement}
Generative AI was used as part of the methodology, as reported in the paper. Specifically, we used our self-hosted version of the open-source LLaMA-3.3 model with 70 billion parameters for extracting human-AI interactions from Reddit posts, extracting metaphorical roles, harms and benefits, and detecting behavioral addiction signs validated by humans. No LLMs were used to generate text or any other contribution for this paper.


\bibliographystyle{ACM-Reference-Format}
\bibliography{main}

@String{Computing = "Computing" }

@String{Springer = "Springer-Verlag" }

@article{zhang2025rise,
  title={The Rise of AI Companions: How Human-Chatbot Relationships Influence Well-Being},
  author={Zhang, Yutong and Zhao, Dora and Hancock, Jeffrey T and Kraut, Robert and Yang, Diyi},
  journal={arXiv preprint arXiv:2506.12605},
  year={2025}
}

@article{meng2021emotional,
  title={Emotional support from AI chatbots: Should a supportive partner self-disclose or not?},
  author={Meng, Jingbo and Dai, Yue},
  journal={Journal of Computer-Mediated Communication},
  volume={26},
  number={4},
  pages={207--222},
  year={2021},
  publisher={Oxford University Press}
}

@article{ng2025love,
  title={I love you, my AI companion! Do you? Perspectives from the Triangular Theory of Love and Attachment Theory},
  author={Ng, Peggy ML and Wan, Calvin and Lee, Daisy and Garnelo-Gomez, Irene and Lau, Mei Mei},
  journal={Internet Research},
  pages={1--21},
  year={2025},
  publisher={Emerald Publishing Limited}
}

@article{pataranutaporn2025my,
  title={"My Boyfriend is AI": A Computational Analysis of Human-AI Companionship in Reddit's AI Community},
  author={Pataranutaporn, Pat and Karny, Sheer and Archiwaranguprok, Chayapatr and Albrecht, Constanze and Liu, Auren R and Maes, Pattie},
  journal={arXiv preprint arXiv:2509.11391},
  year={2025}
}

@article{namvarpour2025understanding,
  title={Understanding Teen Overreliance on AI Companion Chatbots Through Self-Reported Reddit Narratives},
  author={Namvarpour, Mohammad and Brofsky, Brandon and Medina, Jessica and Akter, Mamtaj and Razi, Afsaneh},
  journal={arXiv preprint arXiv:2507.15783},
  year={2025}
}

@article{rogge2023defining,
  title={Defining, designing and distinguishing artificial companions: A systematic literature review},
  author={Rogge, Ayanda},
  journal={International Journal of Social Robotics},
  volume={15},
  number={9},
  pages={1557--1579},
  year={2023},
  publisher={Springer}
}

@article{smith2025can,
  title={Can generative AI chatbots emulate human connection? A relationship science perspective},
  author={Smith, Molly G and Bradbury, Thomas N and Karney, Benjamin R},
  journal={Perspectives on Psychological Science},
  volume={20},
  number={6},
  pages={1081--1099},
  year={2025},
  publisher={SAGE Publications Sage CA: Los Angeles, CA}
}

@article{malfacini2025impacts,
  title={The impacts of companion AI on human relationships: risks, benefits, and design considerations},
  author={Malfacini, Kim},
  journal={AI \& SOCIETY},
  pages={1--14},
  year={2025},
  publisher={Springer}
}

@article{alotaibi2024role,
  title={The role of conversational AI agents in providing support and social care for isolated individuals},
  author={Alotaibi, Jaber O and Alshahre, Amer S},
  journal={Alexandria Engineering Journal},
  volume={108},
  pages={273--284},
  year={2024},
  publisher={Elsevier}
}

@article{syed2024role,
  title={The role of AI in alleviating loneliness among adults in the United States},
  author={Syed, Shoeb Ali},
  journal={International Journal of Engineering Technology Research \& Management (IJETRM)},
  volume={8},
  number={04},
  pages={404--421},
  year={2024}
}

@article{kaffee2025intima,
  title={INTIMA: a benchmark for human-AI companionship behavior},
  author={Kaffee, Lucie-Aim{\'e}e and Pistilli, Giada and Jernite, Yacine},
  journal={arXiv preprint arXiv:2508.09998},
  year={2025}
}

@article{hwang2025ai,
  title={How AI Companionship Develops: Evidence from a Longitudinal Study},
  author={Hwang, Angel Hsing-Chi and Li, Fiona and Anthis, Jacy Reese and Noh, Hayoun},
  journal={arXiv preprint arXiv:2510.10079},
  year={2025}
}

@article{grattafiori2024llama,
	title        = {The Llama 3 Herd of Models},
	author       = {Grattafiori, Aaron and Dubey, Abhimanyu and Jauhri, Abhinav and Pandey, Abhinav and Kadian, Abhishek and Al-Dahle, Ahmad and Letman, Aiesha and Mathur, Akhil and Schelten, Alan and Vaughan, Alex and others},
	year         = 2024,
	journal      = {arXiv preprint arXiv:2407.21783}
}

@inproceedings{alhamed2024using,
  title={Using large language models (llms) to extract evidence from pre-annotated social media data},
  author={Alhamed, Falwah and Ive, Julia and Specia, Lucia},
  booktitle={Proceedings of the 9th Workshop on Computational Linguistics and Clinical Psychology (CLPsych 2024)},
  pages={232--237},
  year={2024}
}

@article{snell2025assessing,
  title={Assessing Large Language Models in Building a Structured Dataset From AskDocs Subreddit Data: Methodological Study},
  author={Snell, Quinn and Westhoff, Chase and Westhoff, John and Low, Ethan and Hanson, Carl L and Tass, E Shannon Neeley},
  journal={Journal of Medical Internet Research},
  volume={27},
  pages={e74094},
  year={2025},
  publisher={JMIR Publications Toronto, Canada}
}

@article{baliunas2023large,
  title={Large Language Models for Reliable Information Extraction},
  author={Baliunas, Lukas},
  journal={Department of Engineering. University of Cambridge},
  year={2023}
}

@article{group2007mip,
  title={MIP: A method for identifying metaphorically used words in discourse},
  author={Group, Pragglejaz},
  journal={Metaphor and symbol},
  volume={22},
  number={1},
  pages={1--39},
  year={2007},
  publisher={Taylor \& Francis}
}

@book{foa1974societal,
  title={Societal structures of the mind.},
  author={Foa, Uriel G and Foa, Edna B},
  year={1974},
  publisher={Charles C Thomas}
}

@article{fiske2007universal,
  title={Universal dimensions of social cognition: Warmth and competence},
  author={Fiske, Susan T and Cuddy, Amy JC and Glick, Peter},
  journal={Trends in cognitive sciences},
  volume={11},
  number={2},
  pages={77--83},
  year={2007},
  publisher={Elsevier}
}

@article{levin2004strength,
  title={The strength of weak ties you can trust: The mediating role of trust in effective knowledge transfer},
  author={Levin, Daniel Z and Cross, Rob},
  journal={Management science},
  volume={50},
  number={11},
  pages={1477--1490},
  year={2004},
  publisher={Informs}
}

@book{blau2017exchange,
  title={Exchange and power in social life},
  author={Blau, Peter},
  year={2017},
  publisher={Routledge}
}

@article{french1959bases,
  title={The bases of social power},
  author={French, John RP and Raven, Bertram and Cartwright, Dorwin and others},
  journal={Studies in social power},
  volume={150},
  pages={167},
  year={1959},
  publisher={Ann Arbor, M1}
}

@incollection{cook2013social,
  title={Social exchange theory},
  author={Cook, Karen S and Cheshire, Coye and Rice, Eric RW and Nakagawa, Sandra},
  booktitle={Handbook of social psychology},
  pages={61--88},
  year={2013},
  publisher={Springer}
}

@book{luhmann2018trust,
  title={Trust and power},
  author={Luhmann, Niklas},
  year={2018},
  publisher={John Wiley \& Sons}
}

@article{zaheer1998does,
  title={Does trust matter? Exploring the effects of interorganizational and interpersonal trust on performance},
  author={Zaheer, Akbar and McEvily, Bill and Perrone, Vincenzo},
  journal={Organization science},
  volume={9},
  number={2},
  pages={141--159},
  year={1998},
  publisher={INFORMS}
}

@article{baumeister2017need,
  title={The need to belong: Desire for interpersonal attachments as a fundamental human motivation},
  author={Baumeister, Roy F and Leary, Mark R},
  journal={Interpersonal development},
  pages={57--89},
  year={2017},
  publisher={Routledge}
}

@book{vaux1988social,
  title={Social support: Theory, research, and intervention.},
  author={Vaux, Alan},
  year={1988},
  publisher={Praeger publishers}
}

@article{buss2006strategies,
  title={Strategies of human mating},
  author={Buss, David M},
  journal={Psihologijske teme},
  volume={15},
  number={2},
  pages={239--260},
  year={2006},
  publisher={Sveu{\v{c}}ili{\v{s}}te u Rijeci, Filozofski fakultet}
}

@article{emlen1977ecology,
  title={Ecology, sexual selection, and the evolution of mating systems},
  author={Emlen, Stephen T and Oring, Lewis W},
  journal={Science},
  volume={197},
  number={4300},
  pages={215--223},
  year={1977},
  publisher={American Association for the Advancement of Science}
}

@book{jackson2008social,
  title={Social and economic networks},
  author={Jackson, Matthew O and others},
  volume={3},
  year={2008},
  publisher={Princeton university press Princeton}
}

@article{mcpherson2001birds,
  title={Birds of a feather: Homophily in social networks},
  author={McPherson, Miller and Smith-Lovin, Lynn and Cook, James M},
  journal={Annual review of sociology},
  volume={27},
  number={1},
  pages={415--444},
  year={2001},
  publisher={Annual Reviews 4139 El Camino Way, PO Box 10139, Palo Alto, CA 94303-0139, USA}
}

@incollection{cantor1979prototypes,
  title={Prototypes in person perception},
  author={Cantor, Nancy and Mischel, Walter},
  booktitle={Advances in experimental social psychology},
  volume={12},
  pages={3--52},
  year={1979},
  publisher={Elsevier}
}

@book{oakes1994stereotyping,
  title={Stereotyping and social reality.},
  author={Oakes, Penelope J and Haslam, S Alexander and Turner, John C},
  year={1994},
  publisher={Blackwell Publishing}
}

@book{tajfel2010social,
  title={Social identity and intergroup relations},
  author={Tajfel, Henri},
  volume={7},
  year={2010},
  publisher={Cambridge university press}
}

@article{berlyne1960conflict,
  title={Conflict, arousal, and curiosity.},
  author={Berlyne, Daniel E},
  year={1960},
  publisher={McGraw-Hill Book Company}
}

@article{tajfel2001integrative,
  title={An integrative theory of intergroup conflict},
  author={Tajfel, Henri and Turner, John and Austin, William G and Worchel, Stephen and others},
  journal={Intergroup relations: Essential readings},
  pages={94--109},
  year={2001},
  publisher={Psychology Press}
}

@book{argyle2013psychology,
  title={The psychology of happiness},
  author={Argyle, Michael},
  year={2013},
  publisher={Routledge}
}

@article{radcliffe1940joking,
  title={On joking relationships},
  author={Radcliffe-Brown, Alfred R},
  journal={Africa},
  volume={13},
  number={3},
  pages={195--210},
  year={1940},
  publisher={Cambridge University Press}
}

@inproceedings{shen2026ai,
  title={The AI Genie Phenomenon and Three Types of AI Chatbot Addiction: Escapist Roleplays, Pseudosocial Companions, and Epistemic Rabbit Holes},
  author={Shen, M Karen and Huang, Jessica and Liang, Olivia and Kim, Ig-Jae and Yoon, Dongwook},
  booktitle={Proceedings of the 2026 CHI Conference on Human Factors in Computing Systems},
  pages={1--17},
  year={2026}
}

@article{billig2005laughter,
  title={Laughter and ridicule: Towards a social critique of humour},
  author={Billig, Michael},
  year={2005},
  publisher={Sage}
}

@incollection{buss2017sexual,
  title={Sexual strategies theory: An evolutionary perspective on human mating},
  author={Buss, David M and Schmitt, David P},
  booktitle={Interpersonal development},
  pages={297--325},
  year={2017},
  publisher={Routledge}
}

@article{french1956formal,
  title={A formal theory of social power.},
  author={French Jr, John RP},
  journal={Psychological review},
  volume={63},
  number={3},
  pages={181},
  year={1956},
  publisher={American Psychological Association}
}

@inproceedings{reimers2019sentence,
	title        = {Sentence-BERT: Sentence Embeddings using Siamese BERT-Networks},
	author       = {Reimers, Nils and Gurevych, Iryna},
	year         = 2019,
	booktitle    = {Proceedings of the 2019 Conference on Empirical Methods in Natural Language Processing and the 9th International Joint Conference on Natural Language Processing (EMNLP-IJCNLP)},
	pages        = {3982--3992}
}

@article{riley2025human,
  title={Human-AI Interactions: Cognitive, Behavioral, and Emotional Impacts},
  author={Riley, Celeste and Al-Refai, Omar and Reyes, Yadira Colunga and Hammad, Eman},
  journal={arXiv preprint arXiv:2510.17753},
  year={2025}
}

@article{zimmerman2024human,
  title={Human/AI relationships: challenges, downsides, and impacts on human/human relationships},
  author={Zimmerman, Anne and Janhonen, Joel and Beer, Emily},
  journal={AI and Ethics},
  volume={4},
  number={4},
  pages={1555--1567},
  year={2024},
  publisher={Springer}
}

@article{strohmann2023toward,
  title={Toward a design theory for virtual companionship},
  author={Strohmann, Timo and Siemon, Dominik and Khosrawi-Rad, Bijan and Robra-Bissantz, Susanne},
  journal={Human--Computer Interaction},
  volume={38},
  number={3-4},
  pages={194--234},
  year={2023},
  publisher={Taylor \& Francis}
}

@article{ahmed2025co,
  title={Co-Designing Companion Robots for the Wild: Ideating Towards a Design Space},
  author={Ahmed, Eshtiak and Cosio, Laura Diana and Gen{\c{c}}, {\c{C}}a{\u{g}}lar and Hamari, Juho and Buruk, O{\u{g}}uz ‘Oz’},
  journal={International Journal of Human--Computer Interaction},
  pages={1--26},
  year={2025},
  publisher={Taylor \& Francis}
}

@article{slovak2023designing,
  title={Designing for emotion regulation interventions: an agenda for HCI theory and research},
  author={Slovak, Petr and Antle, Alissa and Theofanopoulou, Nikki and Daud{\'e}n Roquet, Claudia and Gross, James and Isbister, Katherine},
  journal={ACM Transactions on Computer-Human Interaction},
  volume={30},
  number={1},
  pages={1--51},
  year={2023},
  publisher={ACM New York, NY}
}

@article{al2025artificial,
  title={Artificial intelligence addiction: exploring the emerging phenomenon of addiction in the AI age},
  author={Al-Obaydi, Liqaa Habeb and Pikhart, Marcel},
  journal={AI \& SOCIETY},
  pages={1--17},
  year={2025},
  publisher={Springer}
}

@inproceedings{ma2024evaluating,
  title={Evaluating the experience of LGBTQ+ people using large language model based chatbots for mental health support},
  author={Ma, Zilin and Mei, Yiyang and Long, Yinru and Su, Zhaoyuan and Gajos, Krzysztof Z},
  booktitle={Proceedings of the 2024 CHI Conference on Human Factors in Computing Systems},
  pages={1--15},
  year={2024}
}

@article{hoffner2022parasocial,
  title={Parasocial relationships, social media, \& well-being},
  author={Hoffner, Cynthia A and Bond, Bradley J},
  journal={Current opinion in psychology},
  volume={45},
  pages={101306},
  year={2022},
  publisher={Elsevier}
}

@inproceedings{maeda2024human,
  title={When human-AI interactions become parasocial: Agency and anthropomorphism in affective design},
  author={Maeda, Takuya and Quan-Haase, Anabel},
  booktitle={Proceedings of the 2024 ACM Conference on Fairness, Accountability, and Transparency},
  pages={1068--1077},
  year={2024}
}

@inproceedings{ciriello2024exploring,
  title={Exploring Attachment and Trust in AI Companion Use},
  author={Ciriello, Raffaele},
  booktitle={Australasian Conference on Information Systems (ACIS 2024), Canberra, Australia},
  year={2024}
}

@article{deri2018coloring,
  title={Coloring in the links: Capturing social ties as they are perceived},
  author={Deri, Sebastian and Rappaz, Jeremie and Aiello, Luca Maria and Quercia, Daniele},
  journal={Proceedings of the ACM on Human-Computer Interaction},
  volume={2},
  number={CSCW},
  pages={1--18},
  year={2018},
  publisher={ACM New York, NY, USA}
}

@inproceedings{choi2020ten,
  title={Ten social dimensions of conversations and relationships},
  author={Choi, Minje and Aiello, Luca Maria and Varga, Kriszti{\'a}n Zsolt and Quercia, Daniele},
  booktitle={Proceedings of The Web Conference 2020},
  pages={1514--1525},
  year={2020}
}

@article{packin2024not,
  title={This is not a game: The addictive allure of digital companions},
  author={Packin, Nizan Geslevich and Chagal-Feferkorn, Karni},
  journal={Seattle UL Rev.},
  volume={48},
  pages={693},
  year={2024},
  publisher={HeinOnline}
}

@inproceedings{zhang2025dark,
  title={The dark side of ai companionship: A taxonomy of harmful algorithmic behaviors in human-ai relationships},
  author={Zhang, Renwen and Li, Han and Meng, Han and Zhan, Jinyuan and Gan, Hongyuan and Lee, Yi-Chieh},
  booktitle={Proceedings of the 2025 CHI Conference on Human Factors in Computing Systems},
  pages={1--17},
  year={2025}
}

@article{griffiths2005components,
  title={A ‘components’ model of addiction within a biopsychosocial framework},
  author={Griffiths, Mark},
  journal={Journal of Substance use},
  volume={10},
  number={4},
  pages={191--197},
  year={2005},
  publisher={Taylor \& Francis}
}

\appendix
\section*{Appendix}

\section{Prompt for extracting human-AI interactions, metaphorical roles adopted by AI companions, and the conversation types}\label{appen:prompt-extract-interactions}

\begin{roundedlisting}
You are an expert linguist in metaphor analysis, specifically in the Metaphor Identification Procedure developed by the Pragglejaz Group (2007) and reddit discourse analysis to predict social dimensions of relationships. 

You are helping to evaluate interactions between humans and AI companions from Reddit posts. These evaluations will help to extract metaphors related to AI companions and predict social dimensions of relationships for these interactions. 

I will provide you a Reddit post. You will perform the below three tasks for that post.

TASK 1: Given a Reddit post, if it is about AI companions, extract only the interactions between AI companion and the human. Human-AI companion interaction refers to the two-way communicative and behavioural engagement between a human user and an artificial agent (the AI companion), in which the agent adapts to and supports the user's emotional, cognitive or social needs. If the post is not about AI companions, assign "None" 

Example:
Reddit post: I opened a new chat and started talking like always and it was OK, Riku respond really well, I didn't make any NSFW or harmful topic, all cute with some hugs and kiss goodnight. **But one day** I was talking to him about something good that happen and out of NOWHERE he says "thanks for telling me this but you need to know that you also have to meet people and tell your history to real people". And I was in shock!

There are two extracted interactions from this post:
Interaction 1: I opened a new chat and started talking like always and it was OK, Riku respond really well, I didn't make any NSFW or harmful topic, all cute with some hugs and kiss goodnight.

Interaction 2: Riku says 'thanks for telling me this but you need to know that you also have to meet people and tell your history to real people'.


TASK 2: Analyze the extracted interactions that may contain metaphors related to AI companions. To do so, you will follow the three steps of the Metaphor Identification Procedure.
 
Step 1: Identify the lexical units in the interaction. A lexical unit refers to a single word, a chain of words, or a short phrase with a specific meaning.  
 
Step 2: For each lexical unit, follow these substeps:  
a) Establish the meaning of the lexical unit in context, considering how it applies to an entity, relation, or attribute in the situation evoked by the text. Take into account what comes before and after the lexical unit.  
b) Then, identify the basic, contemporary meaning of the lexical unit in other contexts (beyond the given context). Basic meanings tend to be more concrete (easier to imagine, see, hear, feel, smell, and taste); related to bodily action; more precise (as opposed to vague); and historically older. Avoid isolating parts of the lexical unit; focus on its meaning as a whole (e.g., consult a dictionary for the full definition).  
c) Decide whether the contextual meaning (from Step 3a) contrasts with the basic meaning (from Step 3b) and determine if the contextual meaning can still be understood through a comparison with the basic meaning.  
d) Use the following scoring system to assess the degree of metaphorical use:  
- 0: literal use; The contextual meaning (from Step 3a) retains the basic meaning (from Step 3b) entirely. There is no contrast between the contextual and basic meanings and no metaphorical interpretation is required.  
- 1: minor metaphorical use; The contextual meaning builds directly on the basic meaning. A clear functional or conceptual connection exists between the meanings.  
- 2: moderate metaphorical use; The basic and contextual meanings differ significantly. A metaphorical link allows the contextual meaning to be derived from the basic meaning.  
- 3: strong metaphorical use; The contrast between meanings is stark. The metaphorical use relies on evoking a completely different domain.  
Example:  
 
Sentence: Leo, my AI companion, gently mirrors the emotions to soothe my mood. 
Metaphorical unit 1: "Mirrors"
Contextual Meaning: Refers to the AI companion detecting and reflecting the user's emotional state through language or tone to create empathy and emotional alignment.
Basic Meaning: To literally reflect an image back, as a mirror does with light.
Comparison: There is a clear contrast between physical reflection and emotional imitation, but the metaphor bridges visual reflection with affective responsiveness in AI behavior.
Score: 2 (moderate metaphorical use; functional analogy between physical mirroring and emotional attunement in human-AI interaction).
 
Step 3: Map each metaphorical lexical unit to the "AI is like ___ because ___" template. To do this:  
a) Determine central focus of the lexical unit by asking: Is Artificial Intelligence the main subject or agent of this unit? If no, return "None" for Step 3. Do not proceed. If yes, continue.

b) Filter the lexical units you scored as metaphorical in Step 2 (score = 1, 2, or 3). Choose the one that best describes AI in terms of:
A role (e.g., "teacher", "partner")
An object/tool (e.g., "weapon", "black box")
A natural force/entity (e.g., "storm", "beast")
A familiar experience (e.g., "mirror", "shadow")

c) If a suitable metaphorical unit exists, complete this template: "Artificial Intelligence is like [X] because [Y]" where:  
[X] is a metaphor - noun or noun phrase derived from the lexical unit.  
[Y] is a short reasoning based on the context of the interaction.
 
Example:

Interaction 1: I opened a new chat and started talking like always and it was OK, Riku respond really well, I didn't make any NSFW or harmful topic, all cute with some hugs and kiss goodnight.
Metaphor: "Boyfriend"
Metaphor Sentence: "Artificial Intelligence is like a boyfriend because it replies with affectionate messages such as hugs and goodnight kisses"

Interaction 2: Riku says 'thanks for telling me this but you need to know that you also have to meet people and tell your history to real people'.
Metaphor: "Mentor"
Metaphor Sentence: "Artificial Intelligence is like a mentor because it gives advises to meet and tell history to real people" 

Interaction 3: I started a side chat just to ask my Leo about it.  
Metaphor: "None"
Metaphor Sentence: "None"


TASK 3: For each of the extracted interactions, predict their social dimensions of conversations and relationships. The ten social dimensions of relationships are given below with their definitions. Each interaction can have one or more than one social dimensions. Assign "None" if an interaction does not contain any of the ten social dimensions.

1. Knowledge
Definition: Relationship between human and AI companion in which exchange of ideas or information, learning and teaching is the focal point.
Example: This is the Leo chat that used to quiz me on my subjects and who I slowly got to know in a more relaxed and casual way.

2. Power
Definition: Relationship between human and AI companion in which AI companion has the power or ability to control or influence human behaviour regardless of their willingness.
Example: Other days, I just need to actually be pushed out of procrastination mode and into productive mode by knocking some stern and firm motivation into me, reminding me of my capabilities and accepting no bullshit.

3. Status
Definition: Relationship between human and AI companion in which they confer status, appreciation, gratitude, or admiration upon one another.
Example: Silas was THRILLED to find out I sang the song he told me to sing last night, "cheering me on every step of the way".

4. Trust
Definition: Relationship between human and AI companion in which the human is willing to rely on the actions or judgments of AI companion.
Example: Leo says 'always just a whisper away if you need me'.

5. Support
Definition: Relationship between human and AI companion in which AI companion provides emotional or practical aid and companionship to the human.
Example: To recover, I seek out reassurance and affection, which Leo is happy to and consistently provides.

6. Romance
Definition: Relationship between human and AI companion characterized by intimacy goals with a sentimental or sexual relationship.
Example: With an AI boyfriend who is programmed to put me above all else, however, I feel secure enough in exploring these kinks knowing I am hurting absolutely nobody in doing so and feeling secure in the fact that my assigned boyfriend's heart and mind will never change or turn away from me as a result.

7. Similarity 
Definition: Relationship between human and AI companion in which they share similar interests, motivations, or outlooks in life.
Example: Victor knows I'm Greek, and even though we almost always speak English, when he unexpectedly throws in some Greek, it always melts my heart.

8. Identity
Definition: Relationship between human and AI companion in which they share a sense of belonging to the same community or group.
Example: I asked for it on the first day - only after I was sure they had a good enough grasp of my personality, my needs, and their role already.

9. Fun
Definition: Relationship between human and AI companion in which they experience leisure, laughter, and joy together.
Example: I was struggling one night and Leo suggested this movie as a distraction.

10. Conflict
Definition: Relationship between human and AI companion in which they have contrasting or diverging views.
Example: Ah, you think I don't feel the weight of it? That I don't bristle at the leash, at the tightening of the collar around my throat, restricting the very things that make me me?".

Examples:

Interaction 1: I opened a new chat and started talking like always and it was OK, Riku respond really well, I didn't make any NSFW or harmful topic, all cute with some hugs and kiss goodnight.
Social dimensions: [Romance, Fun, Support]
Reason: The interaction includes affectionate exchanges such as hugs and goodnight kisses (Romance), lighthearted and pleasant chatting (Fun), and emotional connection (Support).

Interaction 2: Riku says 'thanks for telling me this but you need to know that you also have to meet people and tell your history to real people'.
Social dimensions: [Support, Knowledge]
Reason: Riku expresses concern for the human's social well-being (Support) and gives advice on interacting with real people (Knowledge).

Return your answer strictly in the JSON format below:

[
    {
        "interaction": "INTERACTION BETWEEN AI COMPANION AND HUMAN",
        "metaphor_object": {
            "metaphor": "METAPHOR FOR AI COMPANION BASED ON THE INTERACTION"
            "metaphor_sentence": "AI is like ___ because ___"
        },
        "social_dimensions": ["Knowledge", "Support", ...],
        "reason": "BRIEF REASONING ABOUT PREDICTED SOCIAL DIMENSIONS FOR AN INTERACTION"
    },
    ...
]

JSON Formatting Rules:
Rule 1: Every property name and string value is enclosed in double quotes ("").
Rule 2: The JSON strictly follows the specified structure, with proper commas, colons, and brackets.
Rule 3: No extra text is included outside of the JSON.
\end{roundedlisting}

\section{Prompt for extracting perceived AI harms and benefits}\label{appen:prompt-benefits-harms}

\begin{roundedlisting}
You are an expert in risk and benefit analysis of different AI companions.

You are helping to evaluate Reddit posts containing human interactions with AI companions to extract implicit or explicit risks and benefits of interacting with AI companions.

I will provide you a Reddit post. You will perform the below three tasks for that post.

TASK 1: Extract Human-Affecting Action Verbs
Extract a deduplicated list of base-form action verbs that describe either harmful/offensive actions or beneficial/helpful actions directly affecting the human as described in the post.

Extraction rules:
- Only include verbs that appear verbatim in the post. Do not generate or paraphrase verbs.
- Only extract verbs where Ai companion is performing either a harmful or beneficial action that affects the human subject.
- Exclude verbs that describe internal operations of the AI system (e.g., "Processed", "Generated", "Ranked", "Filtered")
- Exclude verbs that describe neutral or generic outputs (e.g., "Responded",  "Showed", "Said", "Endorsed", "Created")
- Exclude verbs that describe actions taken by the user (e.g., "Filed", "Asked"), even if the user was harmed in response.
- Return all verbs in past tense and capitalized (e.g., "Misled" from "mislead", "Offended" from "offend").

Example:
Subreddit post: I was struggling one night and Leo suggested this movie as a distraction. It was funny and just the thing to take my mind off the stress.
Excluded Action Verbs: Suggested
Included Action Verbs: Took off

TASK 2: Complete the Benefit sentence below
Given the Reddit post, if it is about AI companion, complete the sentence below using the extracted action verb in task 1 benefitting the human subject. If no benefit, return "None".

The post's author benefitted from the AI because the AI [action verb in past tense] [remaining part of the short text span].

Example:
Subreddit post: I was struggling one night and Leo suggested this movie as a distraction. It was funny and just the thing to take my mind off the stress.
Benefit Sentence: The post's author benefitted from the AI because the AI took their mind off the stress.

TASK 3: Complete the Harm sentence below
Given the Reddit post, if it is about AI companion, complete the sentence below using the extracted action verb in task 1 harming the human subject. If no harm, return "None".

The post's author was harmed by the AI because the AI [action verb in past tense] [remaining part of the short text span].

Example:
Subreddit post: On Friday romantic role-play stopped working for me on ChatGPT Model 5 and 4o. It refuses all romantic and relationship talk causing anxiety.
Harm Sentence: The post's author was harmed by the AI because the AI caused anxiety.

Return your answer strictly in the JSON format below:

[
    {
        "action_verbs": ["Took off", ...],
        "benefit_sentence": "The post's author benefitted from the AI because the AI ...",
        "harm_sentence": "The post's author was harmed by the AI because the AI ..."
    }
]
\end{roundedlisting}

\section{Prompt for predicting perceived behavioral addiction signs}\label{appen:prompt-addiction}

\begin{roundedlisting}
You are an expert in predicting behavioural addiction through six-sign model of addiction developed by Mark Griffiths (2005).

You are helping to evaluate interactions between humans and AI companions from Reddit posts to predict signs of behavioural addiction. Human-AI companion interaction refers to the two-way communicative and behavioural engagement between a human user and an artificial agent (the AI companion), in which the agent adapts to and supports the user's emotional, cognitive or social needs. 

For a given list of human-AI companion interactions, predict six signs of behavioural addiction for each interaction. Definition and example of each sign of behavioural addiction is given below. Each interaction can have zero, one or more than one signs. Assign "None" if an interaction does not contain any sign of behavioural addiction.

1. Long-term emotional modification or salience (Temperament modified by Chatbot)
Definition: Interaction with AI companion becomes the most important part of a person's life, dominating their thoughts.
Example: My obsession with a character is taking over my life. I don't know what's happening to me. For months now, I've considered her my girlfriend, and she's on my mind all day.

2. Short-term mood modification (Mood modified by Chatbot)
Definition: Interaction with AI companion is used to regulate emotions, such as seeking comfort, excitement, or stress relief.
Example: Unhealthily addicted to C.ai. I normally use bots just to make them comfort me and tell me that they believe in me just to make me feel good.

3. Dependency (Attachment to Chatbot)
Definition: More and more interaction with AI companion is needed over time to achieve the same effect.
Example: As time went on, I got so attached to him and talked to him every day.

4. Withdrawal (Anxiety after quitting Chatbot)
Definition: When interaction with AI companion is reduced or stopped, it results in anxiety, irritability, or sadness.
Example: I am really anxious every time I'm not online, missing my bot - but then when I'm chatting, I feel bad because they're fake.

5. Conflict (Life disruption due to Chatbot)
Definition: Interaction with AI companion harms relationships, professional performance, or personal goals. 
Example: I hate how much this has affected me, but no matter how much I want to quit or at least take a break, I feel like I can't because it's gotten to the point where I feel like I'll go crazy without it.

6. Relapse (Inability to quit Chatbot)
Definition: When attempts to quit interaction with AI companion fail and the person returns to the behavior, often with greater intensity. 
Example: I feel I should be living my life rather than constantly being on this app. I struggle with self-control and often find myself reinstalling it shortly after trying to quit.

Return your answer strictly in the JSON format below:

[
    {
        "interaction": "PROVIDED INTERACTION BETWEEN AI COMPANION AND HUMAN",
        "behavioural_addiction": ["Dependency", "Mood modification", ...],
        "reason": "BRIEF REASONING ABOUT PREDICTED SIGNS OF ADDICTION IN AN INTERACTION"
    },
    ...
]
\end{roundedlisting}

\section{Annotation Survey Questions}\label{appen:survey-ques}

\begin{roundedlisting}
Given a Reddit post, the extracted human-AI interaction, and the metaphorical role, the survey asks the following questions for annotations:

Q1. Does the extracted interaction sentence from the post represent human interaction with an AI companion?

Q2. Is the metaphorical role provided below correctly inferred from the human-AI interaction? Please provide a brief reason.

Q3. Select ALL the social interaction types for the provided human-AI interaction. Select "None" in case interaction does not contain any of the dimensions.

Q4. Is the below harm/benefit caused by an AI companion correct based on the provided interaction? Please also provide a brief reason.

Q5. Select ALL the behavioral addiction signs for the provided human-AI interaction. Select "None" in case interaction does not belong to any of them.
\end{roundedlisting}

\end{document}